\documentclass[fleqn,10pt]{wlscirep}

\usepackage{graphicx}
\usepackage{bm}
\usepackage{epsfig}
\usepackage{graphics,dcolumn,bm,epic, eepic,fleqn,float}
\usepackage{dcolumn}
\usepackage{color}
\usepackage{xcolor}

\title{{On the thermodynamic origin of metabolic scaling}}

\author[1,*]{Fernando J. Ballesteros}
\author[1]{Vicent J. Martinez}
\author[2]{Bartolo Luque}
\author[3]{Lucas Lacasa}
\author[4]{Enric Valor}
\author[5]{Andr\' es Moya}
\affil[1]{Observatori Astron\`omic,  Universitat de Val\`encia, Parque Cient\'ifico de la Universitat de Val\`encia, Paterna (Spain).}
\affil[2]{Departamento de Matem\'atica Aplicada y Estad\'istica \\ ETSI Aeronauticos, Universidad Polit\'ecnica de Madrid, Madrid (Spain).}
\affil[3]{School of Mathematical Sciences, Queen Mary University of London, Mile End Road, London E14NS (UK).}
\affil[4]{Departament de F\'isica de la Terra i Termodin\`amica, Universitat de Val\`encia, Valencia (Spain).}
\affil[5]{Instituto de Biolog\'ia Integrativa de Sistemas, Universitat de Val\`encia-CSIC, Parque Cient\'ifico de la Universitat de Val\`encia, Paterna (Spain)}

\affil[*]{fernando.ballesteros@uv.es}

\begin{abstract}
The origin and shape of metabolic scaling has been controversial since Kleiber found that basal metabolic rate of animals seemed to vary as a power law of their body mass with exponent 3/4, instead of 2/3, as a surface-to-volume argument predicts. The universality of exponent 3/4 -claimed in terms of the fractal properties of the nutrient network- has recently been challenged according to empirical evidence that observed a wealth of robust exponents deviating from 3/4. Here we present a conceptually simple thermodynamic framework, {where the dependence of metabolic rate with body mass emerges from a trade-off between the energy dissipated as heat and the energy efficiently used by the organism to maintain its metabolism. This balance tunes the shape of an additive model from which different effective scalings can be recovered as particular cases,} thereby reconciling previously inconsistent empirical evidence in mammals, birds, insects and even plants under a unified framework. This model is biologically motivated, fits remarkably well the data, and also explains additional features such as the relation between energy lost as heat and mass, the role and influence of different climatic environments {or the difference found between endotherms and ectotherms.}
\end{abstract}
\begin{document}

\flushbottom
\maketitle
%
%
\thispagestyle{empty}

\section*{Introduction}
The basal metabolic rate $B$ (kJ/h) is the minimum energy expended daily by an animal in thermoneutral conditions to keep its metabolism at work {(for ectotherms that lack a thermoneutral zone, an alternative concept such as resting or standard metabolic rate at a given temperature is used)}. As early as in 1839, Sarrus and Rameaux \cite{seminal} proposed that metabolic rates might depend on heat dissipation (Fourier's law) and therefore increase with surface area, something originally checked in dogs by Rubner in {1883} \cite{seminal2}.\\ 
{Originally Krogh (1916, \cite{Krogh}) and more popularly Kleiber (1932, \cite{Kleiber})} empirically observed  that, indeed, a simple and robust allometric scaling between $B$ and the animal mass $M$ could account for most of the metabolic rate variability, $B\sim M^{\alpha}$. However, he found that $\alpha=3/4$, instead of $\alpha=2/3$ that results of heat dissipation according to a simple surface-to-volume argument. Since then, extensive data have been collected, encompassing a fervent debate on the origin and concrete shape of the so called Kleiber's law. While some of the empirical works seem to comply better to $\alpha=2/3$ \cite{Heusner, White, Speakman}, a great majority took for granted a 3/4 power law \cite{Bartels, Feldman, Savage}, raising it to the level of \textit{central paradigm} in comparative physiology \cite{review2005}. This scaling was subsequently elegantly explained by space-filling fractal nutrient distribution network models \cite{West, West2, Banavar2, Banavar} (with possible deviations for small masses due to finite size effects), thus apparently closing the debate on its origin. However, additional statistical evidence challenges the validity of $\alpha=3/4$ \cite{Dodds, Kozlowski, Kolokotrones, Chown, McNab, McNab2009, White2, fito} (see also \cite{PLOS} for a recent experimental study of the fractal exponents in human vascular networks). For instance Dodds et al. \cite{Dodds} found that, for masses under 10 kg, a 2/3 exponent gives a better fit, while 3/4 fits better the whole range. In the same vein, Kolokotrones et al. \cite{Kolokotrones}, after fitting the encyclopaedic dataset of basal metabolic rates for mammalians compiled by McNab \cite{McNab}, concluded that the scaling law was not after all a pure power law but had curvature in double logarithmic scales, giving an heuristic explanation as to why different exponents could be fitted depending on the range of masses considered (but see \cite{mackay}). Strong evidence of curvilinearity in the log-log relationship is also reported by Clarke et al. \cite{clarke} who introduced the body temperature to mitigate the effect, {and actually curvilinearity in log-log plots was already suggested by Hayssen et al back in 1985 \cite{Hayssen}}. Other views are indeed  more skeptic about the reality of allometric scaling\cite{Hulbert14}.

\noindent In the last decades, a large number of theories of different garment and degrees of formality have been proposed to justify the occurrence of particular scaling forms \cite{White2, Glazier2014}, organized into four major brands (surface area, resource transport, system composition, and resource demand models) by Glazier \cite{Glazier2014}. For instance, the $\alpha=3/4$ theories relate to the geometry of nutrient supply networks \cite{West, West2, Banavar2} or general geometrical arguments \cite{Banavar} whereas some advocates of the $\alpha=2/3$ include mainly heat dissipation \cite{Heusner, White, Speakman}. To name a few other approaches, in the quantum metabolism (QM) model \cite{Agutter} power laws with varying pre-factors are found. Similarly, {the metabolic level boundaries hypothesis (MLBH) \cite{Glazier} is an important conceptual framework that suggests that the power law exponent of metabolic scaling relationships should indeed vary between 2/3 and 1 according to the particular `metabolic level' (activity level of the organism or metabolic intensity).} The allometric cascade model \cite{Darveau} on the other hand deals with a linear combination of innumerable allometric components, one for each different tissue, and ends with a complex formula depending on too many free parameters to fit. The three-compartment model \cite{Painter} deals with a linear combination of three allometric components that model different classes of organs and tissues. The Dynamic Energy Budget (DEB) \cite{Kooijman} comprises a weighted sum of four processes: assimilation, maintenance, growth, and maturation, the first one with exponent $2/3$ as Kooijman assumes that surfaces assimilating the incomes (oxygen, food) scale as $2/3$ (but West et al. \cite{West} {showed that some respiratory variables in lungs} scale rather as $3/4$). The effect of cell number and size  in body mass has also been claimed to be responsible for different allometric scalings \cite{K}. The abovementioned MLBH \cite{Glazier14} justifies the diversity of metabolic scaling relationships within physical limits related to body volume and surface area. Finally, gravity has been also considered as part of the problem \cite{Economos}, due to response of body mass to gravity, which scales linearly with mass (see \cite{review2005, White2, Glazier2014} for reviews).

Not all of the mechanistic explanations for metabolic scaling are necessarily exclusive, and quite probably many of these remain valid on appropriately defined limits. Integration of those under a more parsimonious hallmark is, however, certainly needed \cite{White2}.
In this work we aim at proposing such a hallmark, with the hope that it might help to reconcile several empirical and statistical results. Not all of the organism's energy income is wasted as heat, and we find that this simple thermodynamic balance allows us to explain in a quantitative way the correct allometric curves for mammals (in different environments), birds, and insects, to account for other biological features such as the relation between energy lost as heat and mass, as well as to extend the analysis to plants.

\section*{The model}
In this paper we advance a simple energy balance model that aims at unifying and reconciling previous models and results. {As already discussed, numerous classes of metabolic models exist with a wide variety of complexities. We here focus on a remarkably simple class of model combining allometric and isometric terms. This model class was first
considered by Swan \cite{Swan} and studied by Yates \cite{Yates}. We will show that our novel interpretation both provides a bottom-up argument for the relevance of this model and shows
that its parameters are physically interpretable. We will show that,
despite being substantially more parsimonious than more complex models discussed above, this model explains well a wealth of data, admits ready physical interpretation, and can explain environmental dependence of metabolic scaling.}\\
To be more concrete, we argue that the trade-off between the energy dissipated as heat and the energy efficiently used by the organism to keep it alive results in a model for the dependence between $B$ and $M$ with an isometric (proportional to $M$) and an allometric (proportional to $M^{2/3}$) term, balanced respectively by prefactors $k$ (with units kJ/hg) and $k'$ (with units kJ/hg$^{2/3}$) called Meeh factors \cite{Meeh}, which have clear biological meaning and can thus be estimated empirically. This balance complies with an effective (apparent) pure power law in a double logarithmic plot, with varying exponent in the range $[2/3-1]$ as proposed by Glazier \cite{isometric, fito}. As shown later, we will be capable of recovering the correct (non)scaling form in mammals, birds and insects datasets, and explain why the apparent power law scalings show different exponents in all these cases, as well as extending the theory to plants. Furthermore, we will also predict an estimation for the energy conversion efficiency of mammals which agrees with independent considerations based on oxydative phosphorylation in mitochondria. We will also predict the different metabolic scaling shapes occurring in different environments (polar and hot desert mammals).

\noindent {A priori, the energy intake is typically proportional to the number of cells in the organism, hence grows isometrically with body mass $M$. Then, the first step is to recall that part of such energy is converted into work and used in a plethora of different metabolic and physiological processes, including the synthesis of ATP and proteins, cellular division, muscle contraction etc: it keeps the animal alive. These processes are thermodynamically inefficient, hence part of the energy consumed is dissipated as heat. Such dissipation is always present, and in the case of endotherms such dissipation is complemented with an additional amount due to internal heat production. The key question then is to assess how efficient such energy conversion is. Let us consider then two extreme (unrealistic) situations: in one end, suppose that the process is totally inefficient, i.e. zero work conversion and all energy dissipated as heat. The organism in that case would be a simple (dead) heater radiating as a black body, using energy intake only to keep itself warm. As this energy would be subsequently dissipated through the organism surface, thermoregulation would thus put an upper bound for the amount of energy that can be consumed, which according to simple surface-to-volume arguments is at most balanced with heat dissipation and then scales allometrically proportional to $M^{2/3}$: this would yield an effective ``metabolic rate" $B = k'M^{2/3}$ for some constant $k'$. At the other extreme, we consider an ideal situation of perfect energy conversion efficiency, where all energy consumed would be efficiently converted into work and the whole body would be a sink of energy with no heat losses. Since energy is originally consumed isometrically, without any further geometric restrictions on heat dissipation, the energy spent would also scale isometrically, hence $B = k M$ for some constant $k$.}\\ 
Now, our contention is that living organisms interpolate between these extrema: they are neither dead heaters nor optimal energy sinks, but lie somewhere in between, as any thermodynamic system operating away from equilibrium. As such, these simple thermodynamic arguments suggest an effective model where, if $f$ is the fraction of the energy income that is used ``efficiently" by the cells to keep their metabolism working, 
and $1-f$ is the fraction of the energy lost as heat, there must be a balance between the isometric and the allometric term, as both mechanisms are present simultaneously. In principle, one can balance out these two terms by two generic weights labeled $w$ and $w'$ respectively, in such a way that the basal metabolic rate would comply with
\begin{equation}
B= w k M+w' k' M^{2/3}.
\label{GBMR}
\end{equation}
Note however that $w$ and $w'$ are not independent, simply because the two associated mechanisms use the same income energy. In other words, there is also a trade-off between them: if too much energy dissipates as heat, the organism keeps little energy for the metabolism. Moreover, these weights are in fact functions of $f$: $w(f)$, $w'(f)$ and have to fulfill the following constraints:
\begin{enumerate}
\item
If $f=1$ (no heat losses) then $w=1$ and $w'=0$.
\item
If $f=0$ (only heat losses) then $w=0$ and $w'=1$. 
\item
Both weights $w$ and $w'$ range from 0 to 1, as $f$ does. 
\end{enumerate}
The simplest choices that fulfill these three requirements are $w(f)=f$, $w'(f)=1-f$, which corresponds in fact to use the factor $f$ itself as the weight of each process (although other possible more complex relations could be considered). Therefore, with this hypothesis, Eq. \ref{GBMR} becomes
\begin{equation}
B = f k M + (1-f) k'M^{2/3}. 
\label{BMR}
\end{equation}

\noindent {Four simple observations are in order: } first, eq. \ref{BMR} is not a pure power law but the linear combination of two, with exponents 1 and 2/3 respectively. It is well known \cite{White2, Yates} that in a double logarithmic plot, this kind of equation yields a curved graph with convex curvature, in good agreement with the findings of Kolokotrones et al.\cite{Kolokotrones}. Second, for small values of $M$ {(e.g. for mammals of small mass)}, this equation approximates very well to a power law with exponent 2/3, and is therefore in accordance with recent results by Dodds et al \cite{Dodds}. Third, for a large range of masses, this equation approximates to an {\it apparent} pure power law with an {\it effective} exponent that can range between 2/3 and 1, in good agreement with empirical evidence (entropic considerations prevent the linear asymptotic regime to appear empirically, see however \cite{isometric}). {Fourth, in the case of ectotherms the allometric term is only associated to heat dissipated due to metabolic and physiological processes and not also due to thermoregulation as for endotherms, and therefore in that case we expect the pre-factor of the allometric term to be much smaller than in the case of endotherms, what would yield a larger apparent exponent if fitted to a single power law (see the next sections for validation).}\\ 
According to Glazier classification of metabolic scaling models \cite{isometric}, our model can be classified within the type III family: a shift for nearly isometric to negatively allometric behavior.\\

\noindent The parameter $f$ and the Meeh pre-factors $k$ and $k'$ have biological meaning and therefore can be measured experimentally (this will be discussed later). However at this point we consider them simply as constants, and since they are independent of $M$ we can reabsorb them by defining $a = fk$ and $b = (1-f)k'$. Therefore our thermodynamically grounded model eq. \ref{BMR} reduces to $B= aM + bM^{2/3}$, which is now a statistical model with only two fitting parameters that can be {\it fitted} to available databases. Note at this point that the use of power law functions as well as sums of power laws is not a new idea. Actually the combination of an isometric term (proportional to $M$) and an allometric term (proportional to $M^{2/3}$) is also part of the  DEB theory \cite{Kooijman, Sousa}, found following different arguments that the ones used here (although in the DEB theory this only applies for intra-species relationship). As Kooijman \cite{Kooijman} states, the intra- and the inter-specific scaling are numerically (although not formally) very similar, but in the first case the scaling responds to the reducing contribution of growth to respiration (obtaining the aforementioned $aM+bM^{2/3}$), while in the second case to the increasing contribution of reserves to body weight (yielding $[aM+bM^{2/3}]/[c+M^{1/3}]$).
A more detailed analysis of the different mechanistic explanations that could account for such a statistical model (e.g. DEB, MLBH) is beyond the scope of this work, and we refer the interested reader to \cite{Sousa08, Sousa10, Nisbet12} and \cite{Glazier14}. We now will proceed to fit our statistical model to experimental data.\\

\begin{figure}
\centering
\includegraphics[width=0.7\columnwidth]{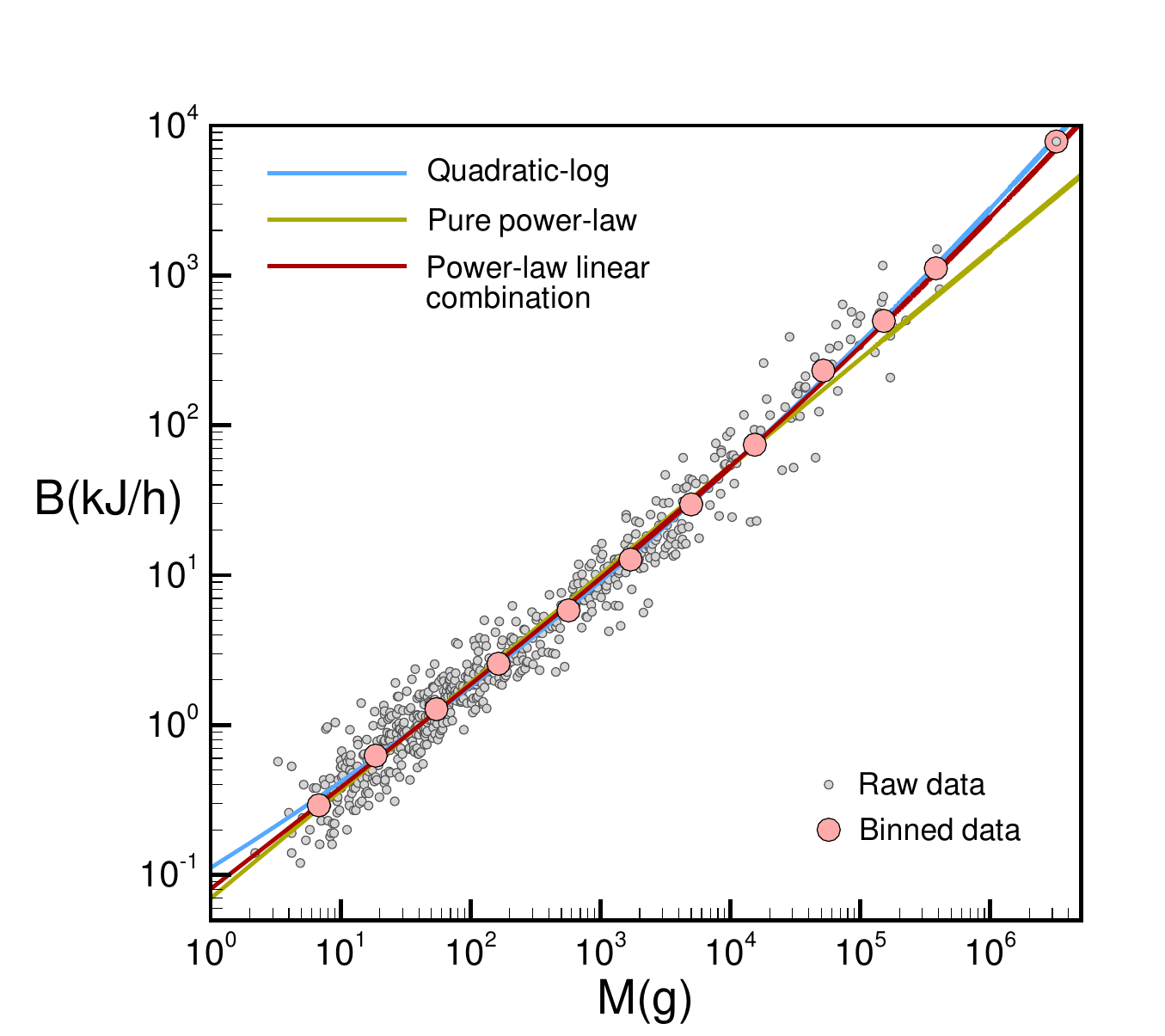}
\caption{{\bf Basal metabolic rate for mammals.} Gray dots: basal metabolic rate data for mammals compiled by McNab \cite{McNab}. Red line: fitting of our theory to the data (see table I for statistical tests). Blue line: Kolokotrones et al. statistical model \cite{Kolokotrones}. Green line: fitting to a pure power law. We also include a logarithmic binning of the data (pink dots) where the curvature is better appreciated. These binned points have been included as a guide to the eyes to enhance the curvature of data, but fits have been performed using the raw data. The size of the points correspond to the one sigma dispersion of the residuals respect to our model for the whole set of data.}
\label{fig_1}
\end{figure}

\noindent {\bf Fitting the model in mammals. }
As can be seen in Fig. \ref{fig_1} and table I, this model fits exceptionally well the collection of $N=637$ mammal basal metabolic rates recently compiled by McNab \cite{McNab}. Statistically, its fitting is as good as the exotic quadratic function $\log B = \beta_0 + \beta_1 \log M + \beta_2(\log M)^2$ proposed by Kolokotrones et al \cite{Kolokotrones} to quantify the curvature underlying metabolic scaling, but has only two free parameters instead of three and, more importantly, is thermodynamically justified. Interestingly, if the exponents were left as free parameters, the best fit would indeed give 1 and 2/3 within a 0.5$\%$ error. Fitting values are $a = fk = 0.0016$ for the {pre-factor associated to efficient energy-work conversion isometric term} and $b =  (1-f)k'=0.079$ for the {pre-factor associated to heat dissipation}, while a pure power law can be approximated with an effective exponent $\alpha\approx0.72$ (see table I).\\
Once we have shown that the model fits exceptionally well the data {(outperforming the pure power law model through a model comparison based on Akaike Information Criterion)}, in what follows we go back to the original model eq. \ref{BMR} and focus on the biological variables $k,k'$ and $f$; we will advance a formula for $k'$ and accordingly predict an estimation for $f$, which we will show to be on good quantitative agreement with independent empirical evidence.\\

\noindent {\bf Estimating $k'$ and $f$. } Note at this point that $b$ and $k'$ should be of the same order of magnitude and $k'> b$ for our model to be consistent. Let us now estimate $k'$. For a pure heater of mass $M$ and density $\rho$  at constant temperature, heat generated inside its volume $V$ is balanced with the heat lost through its surface area $A$. The 'basal metabolic rate' of the heater can then be defined as the total heat loss $Q=qA=k'M^{2/3}$, where $q$ is the energy loss per time and area units. Now, a simple dimensional analysis yields $A = [d/\rho^{2/3}]M^{2/3}$ , where $d$ is a dimensionless number depending on the geometry of the body --$d=6, 4.83$ or $7.2$ for a cube, a sphere and a tetrahedron respectively-- {(note that body shape considerations have been reported to play a role in other works\cite{isometric})}. 
Now, $q$ can be further separated in several components according to the different physical mechanisms that yield heat dissipation.
If we only consider convection $q_C$ and radiation $q_R$ as sources of heat losses (this being a fair approximation under the conditions of basal metabolic rate measurement,
where evaporation through transpiration or respiration is not relevant for most of the mammals), then $q = q_C + q_R$. The first summand $q_C = h_C \Delta T$, where $h_C$ is the convective heat transfer coefficient (which for still air $h_C$ ranges between 3-4 $W/m^2K$ \cite{Blaxter}) and $\Delta T=T_s-T_e$ is the difference between the surface temperature of the mammal ($T_s$) and the environment ($T_e$). Considering on the other hand radiation losses, note that animals radiate heat similar to a black body in infrared wavelengths. Therefore, one can use Stefan-Boltzmann law for black bodies such that $q_R  = \sigma(T_s^4- T_e^4) \approx 4\sigma ((T_e+T_s)/2)^3 \Delta T := h_R \Delta T$, where $\sigma\approx 5.67\cdot10^{-8}Wm^{-2}K^{-4}$ in SI units. Altogether, as $k'M^{2/3}=qA$, solving for $k'$ we end up with a general expression $k'=d(h_R+h_C)\Delta T\rho^{-2/3}$ where, remarkably, all parameters are now empirically observable.

\noindent As a rough approximation, we can now estimate $k'$ by taking average values for all the parameters: $d \approx 6$, a water-like density $\rho \approx 1000 \ kg/m^3$, and $h_C\approx 3.5$. According to Mortola \cite{Mortola} a good average for mammals is $\Delta T \sim 5^\circ C$.
Averaging for $T_e$ ranging between $15^\circ C$ and $30^\circ C$, and for $T_s$ being 1 to 10 degrees higher than $T_e$, $h_R$ gives values between 5.5 and 6.5, thus we take $h_R \approx 6$ as an average.  This yields $k'\approx 3$ in SI units, for which  
$Q(W) = k' M^{2/3} \approx 3M^{2/3}(kg)$. Transforming into appropriate units 
$Q(kJ/h) \approx 0.1 M^{2/3}(g)$, i.e. $k'\approx 0.1$. Note that this estimation depends on several empirical variables that show variability, so this number should be taken with caution (performing a parametric analysis of $k'$ for a range of plausible values: $d\in[5,7],\ h_C\in[3,4], T_e\in[280,300],\ T_s\in[T_e+1,T_e+10]$ we find however $\langle k' \rangle \approx 0.1$ so the approximation is robust). Note that $k'>b$ but of similar order of magnitude, as previously required. Since $b= k'(1-f)\approx 0.079$, our simple approximation predicts $f\approx 0.21$, and the efficient energy metabolic consumption per mass unit  $k = a/f\approx 0.0076 $kJ/hg. Now, whereas (to the best of our knowledge) there are no accurate direct experimental estimates for $f$ in the literature, our prediction can still be tested against experimental evidence in the following terms. The principal sources of heat in the organism of mammals are the synthesis of ATP through the oxidation of nutrients \cite{Rhoades}, and the subsequent use of this ATP as energy source for other biological reactions. Animal ATP is generated inside mitochondria mainly via oxidative phosphorylation, or cellular respiration. It has been estimated \cite{Chen} that this pathway actually produces more that 90\% of ATP. Furthermore, note that glycolisis is discarded here as this alternative pathway is only significant under low levels of oxygen and other circumstance which do not typically hold under the comfortable conditions of basal metabolic rate measurement. Now, the oxidation of glucose produces the energy to create ATP from ADP, however, only about 42$\%$ \cite{Rhoades} of the energy stored in glucose is captured in ATP (the other 58$\%$ being converted into heat), thus $f\leq0.42$. ATP is subsequently used to fuel a plethora of biological reactions, and rough estimations \cite{Barasi} quantify that only about 50$\%$ of this energy is actually used (ATP hydrolysis is usually higher than the energy necessary to carry subsequent biological reactions), and the rest contributes to heat losses. This gives a (totally independent) empirical estimation $f \approx 0.21$, which remarkably matches our prediction.
{Incidentally, note that a recent study \cite{iain} elaborates on models similar to eq.\ref{BMR} at the cellular level to explain observed scaling
relationships between mitochondrial populations, cell growth, and cell
survival}.\\

\begin{figure}
\centering
\includegraphics[width=0.7\columnwidth]{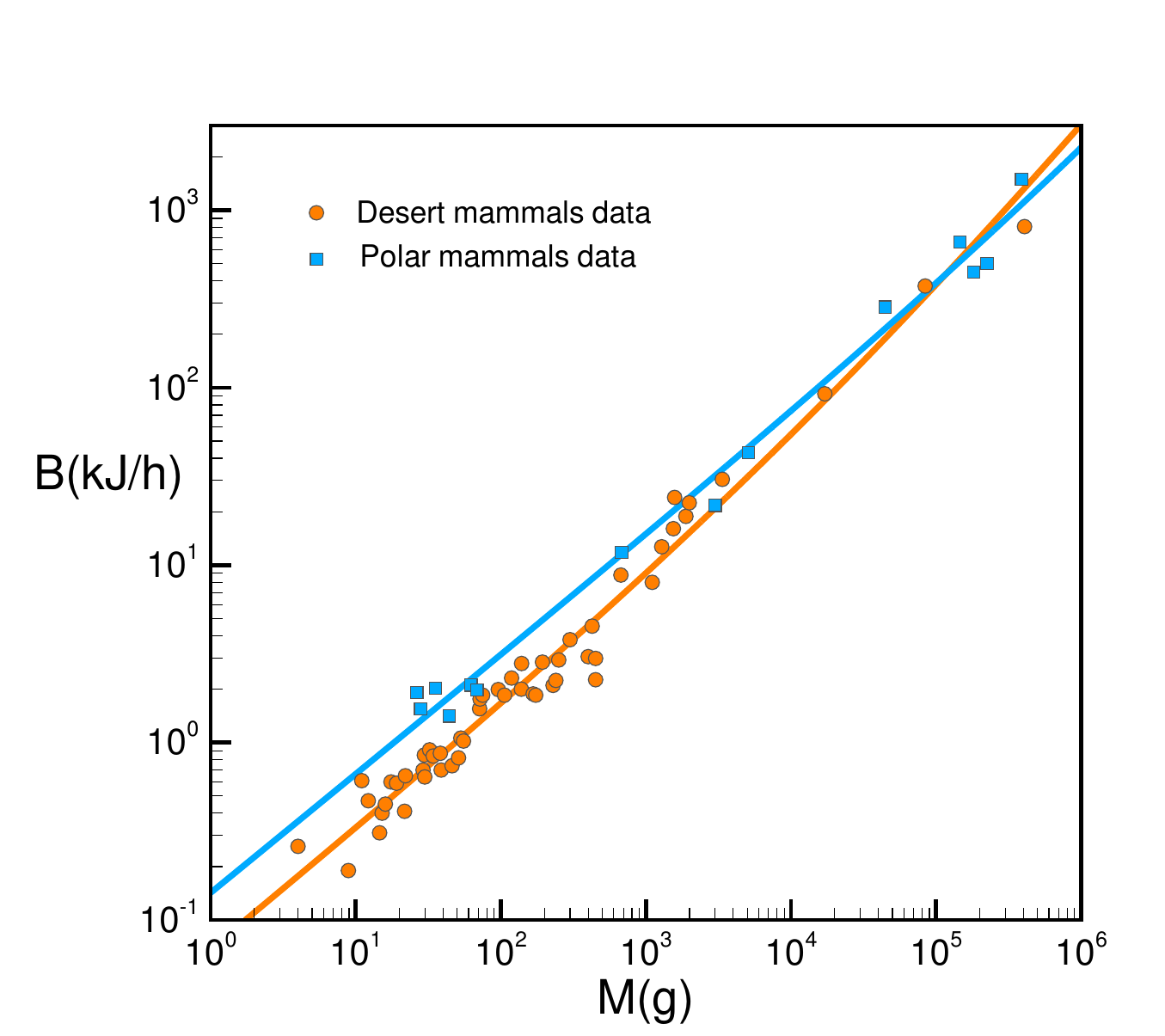}
\caption{{\bf Polar vs desert mammals.} Orange dots: subset from McNab's data \cite{McNab} corresponding to hot desert mammals. Blue squares: subset corresponding to polar mammals. Orange and blue lines: fitting to the model (see table I for statistical tests).}
\label{Fig2}
\end{figure}

\noindent {\bf Varying climatic conditions: Polar vs hot desert mammals. }
The universality of Kleiber's law ultimately stems from the universality of cell's energy source. Consequently, the predicted values for $f$ should be considered a reasonable average value for all mammals. Interestingly enough, mitochondria evidence slight adaptations for animals living in cold and hot environments. For example, for polar mammals, the concentration of thermogenin inside the mitochondria is unusually high (constituting up to 15$\%$ of the total protein in the inner mitochondrial membrane \cite{Lodish}). Thermogenin actually uncouples oxidative phosphorylation from ATP synthesis, causing all energy released by the oxidation of glucose to be released as heat, without creating ATP, hence warming up the animal: this and other similar uncoupling proteins are a way to effectively decrease energy conversion efficiency $f$ in mitochondria.
According to our theory, the climatic adaptations in mitochondrial energetic efficiency should cluster polar and desert mammals,  have a net effect in the respective values of $f$, and thus in the apparent exponent of a pure power law fitting. In particular, the ratio $b/a\propto(1-f)/f$ increases as $f$ decreases, i.e. polar mammals with lower (mitochondrial) energy conversion efficiency should have larger $b/a$ ratios than hot desert ones, and hence lower effective exponents (closer to 2/3) in a pure power law fit according to eq. \ref{BMR}. To test this prediction, we have extracted all polar and hot desert mammals from McNab's dataset, and plotted their basal rates in figure \ref{Fig2} (blue points for polar environments, orange points for hot deserts), along with a fit to the model 
(of course parameter values from these fits will be different than for the whole set, as the whole set shows an average behavior for the whole mammals, with higher dispersion; this is confirmed by the fact that scatter in these subsets respect to their fitting line is smaller than in the whole set). Remarkably, both subsets are  clustered, with polar mammals having on average larger metabolic rates than desert ones for a given mass, in agreement with the results presented by Lovegrove \cite{Lovegrove00, Lovegrove03}. Fristoe et al. \cite{fristoe} shows that shifts in the basal metabolic rate help both birds and mammals to adapt to different environmental temperature regimes.  Both $b/a$ and the effective exponents agree with the predictions of our theory. Interestingly, the effective slope for polars is $\alpha\approx0.69$, a value which is closer to $2/3$, the expected one for pure heaters (see table I for the fitting details). Accordingly, the well-known tendency of polar mammals to be larger than desert ones can be justified in terms of the aforementioned considerations (as polar mammals tend to be more energetically inefficient, they need to be larger to reduce the impact of heat dissipation).\\

\noindent  {From a statistical point of view, one can design a simple statistical experiment which can help to further confirm that predictions from our model are genuine and our model reflects some true underlying effects. The experiment consists in considering the subset of cold and desert mammals altogether (a total of $N=113$ species) and make a model selection for three statistical models: a pure power law (M1), our model (M2) and an hybrid model which fits two versions of Eq.\ref{BMR}: one for desert mammals $a_hM + b_hM^{2/3}$ and another one for cold mammals $a_cM + b_cM^{2/3}$. AIC for M1, M2, and M3 respectively are -233, -245 and -277. Note that M2 outperforms M1 for this subset. More importantly, if clustering of polar and desert mammals was an artifact, then M3 shouldn't outperform M2. However we find much lower AIC in the last case: the relative likelihood of M3 with respect to M2 is approx $\exp(16)\approx9 \cdot 10^6$, providing a compelling statement
that the model reflects some true underlying effects.}\\

\noindent Finally, note that while climatic conditions might have an effect on $k'$ (for instance, differences between skin and environment temperatures play a role in the computation of the radiative source $q_R$), the constant related to efficient energy conversion should be considered similar for the whole set of mammals. If this hypothesis is correct, from Fig. \ref{Fig2} it would follow that $f\approx 0.14$ for polar animals and $f\approx 0.4$ for desert ones, while $k'=0.16$ and $0.109$ respectively. These new predictions await for experimental confirmation.\\

\noindent As $f$ is the fraction that does not appear as heat, one would therefore expect to see differences between direct calorimetry (measuring heat production) and indirect calorimetry (oxygen consumption).  These quantities cannot coincide, otherwise all energy consumed would be lost as heat and this is not possible since a fraction of the chemical energy ingested by the organism must be used for cellular work, and for building blocks for storage, of for growth of the somatic body and for reproductive material.  This is precisely what $f$ quantifies. As we have already mentioned, there are not many reliable measurements of this quantity in the literature. It is clear that in rapidly growing embryos or organisms differences between direct and indirect calorimetry have to be more evident. In fact, Zotin \cite{Zotin} shows several examples of this kind, allowing us to estimate  $f=(Q_{O_2} - Q)/Q_{O_2}$, where $Q_{O_2}$ is the consumed energy  measured by the oxygen consumption (indirect calorimetry)  and $Q$ is the consumed thermal energy measured by the heat lost (direct calorimetry). We can see in the results shown in Zotin  \cite{Zotin} (in particular, Figs. 3.26, 3.38, and 3.29, and tables 3.7 and 3.8), that values of $f$ oscillate between 0.13 and 0.25, again in agreement with our predictions. Moreover, for human adults at rest, we see that $f$ takes a value of about  $\approx 0.15$ (Fig. 3.32 in \cite{Zotin}).\\

\noindent So far our analysis dealt with mammals.
In what follows, we extend this analysis to birds, insects and plants. These are smaller databases than the ones used for mammals which however are large enough for accurate statistical analysis.
We will show that while in these cases a pure power law model provides reasonably similar statistical results than our proposal, the effective exponent found varies from case to case, thus one would need individual ad hoc theories that could explain the particular effective exponent for each case. On the other hand, all the results indeed comply with a combination of isometric and allometric scalings of the shape of equation \ref{BMR}, with varying pre-factors.\\

\begin{figure}
\centering
\includegraphics[width=1.0\columnwidth]{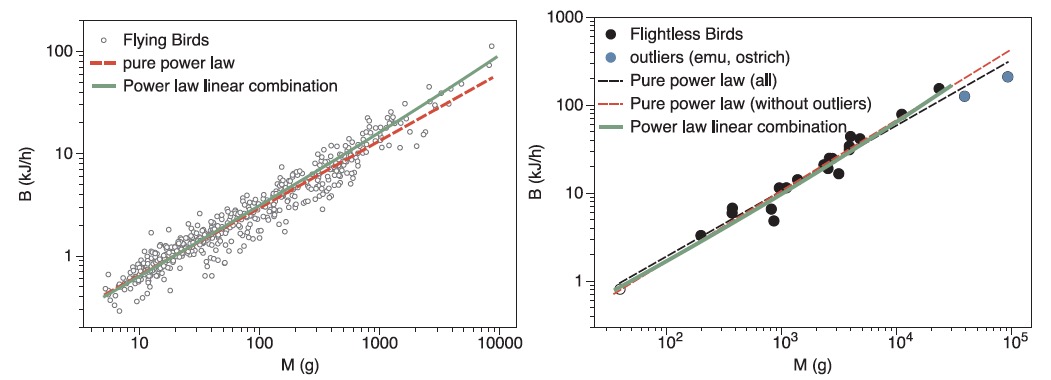}
\caption{{\bf Basal metabolic rate for birds.} Data have been drawn from McNab \cite{McNab2009}, and split into flying species (503 data, left panel) and flightless ones (22 species, right panel. Both apparent exponents (dashed lines) differ $\alpha\approx 0.66$ for flying species, and $\alpha\approx \{0.74,0.8\}$ (depending if we consider outliers  ostriches \cite{outliers1} and emus \cite{outliers2} in the fitting) for flightless ones. The model fit is shown in solid green line (see table I for statistical tests).
}
\label{Fig_p}
\end{figure}

\noindent {\bf Extension to birds and insects. }
As an extension, we first make use of McNab's collection of bird's metabolic rates \cite{McNab2009} (more than 500 species) and Chown et al's insect database \cite{Chown} (more than 300 species). 
In the case of birds, we further split the analysis into flying and flightless species, and plot their metabolic rates in both panels of fig. \ref{Fig_p}. For the case of flying birds (503 species), the apparent power law exponent is $\alpha\approx 0.657$ -deviating from the theoretical prediction 3/4-, but we can see that eq. \ref{BMR} fits reasonably better the whole range. 
For flightless species the dataset is much smaller (22 species). Within this category, note that the largest species (emu and ostriches) are known to have abnormally low metabolic rates \cite{outliers1,outliers2}. In fact, the fitted apparent exponent $\alpha$ varies between 0.74 and 0.8 if these species are removed. In this latter case, no strong differences are found between the pure power law and equation \ref{BMR} (if no splitting between flying and flightless birds is performed, results are very similar to the flying case, as flightless birds are much less common). Note that flying birds tend to have larger values of $B$ than mammals and comparatively behave closer to the 'heater' limit $\alpha=2/3$. On the other hand, flightless birds cluster towards lower metabolic rates than flying ones, and behave closer to mammals (as a matter of fact, the rates for flightless birds are compatible with the curve found for mammals).\\

\begin{figure}
\centering
\includegraphics[width=0.7\columnwidth]{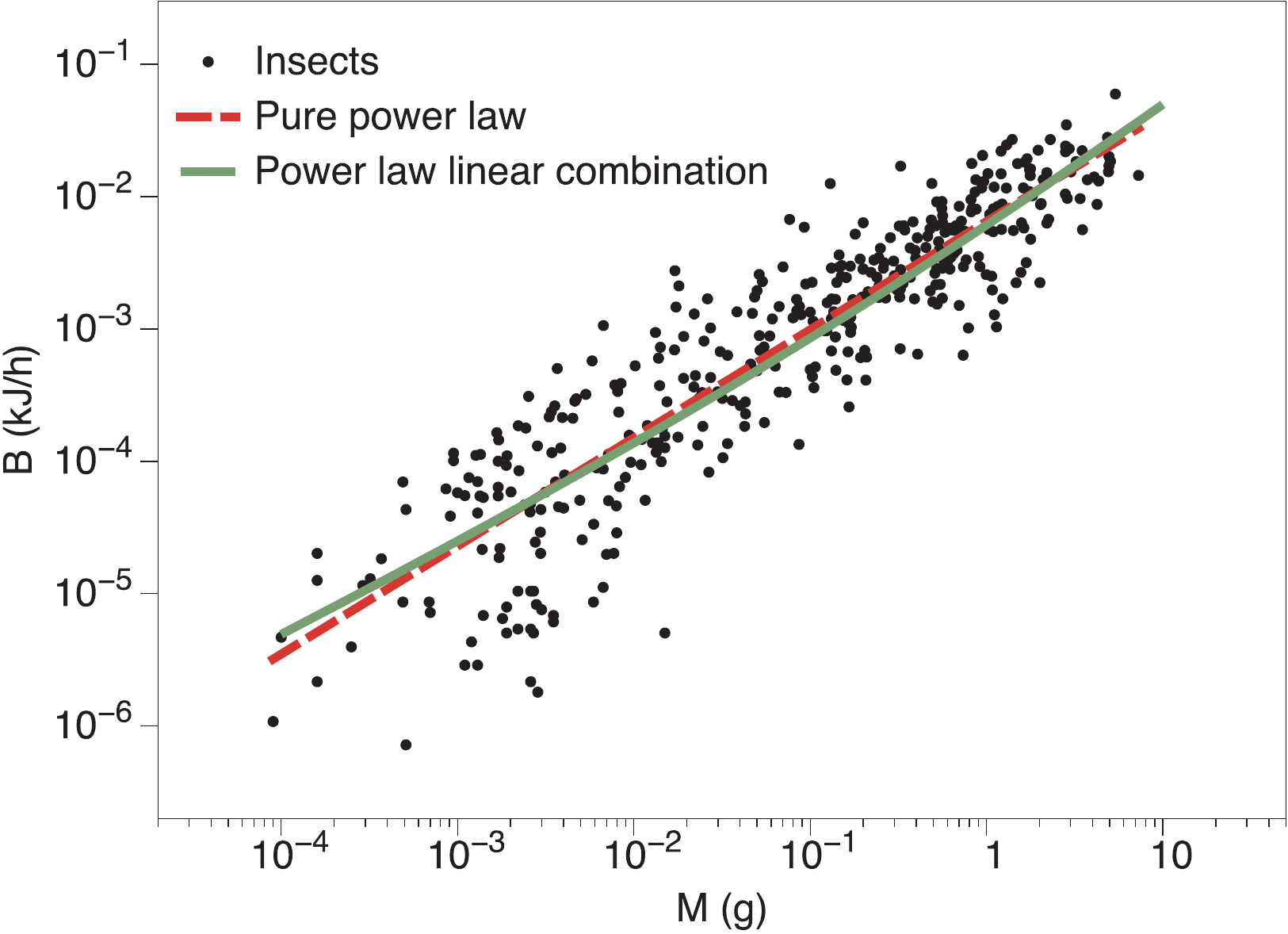}
\caption{{\bf Metabolic rate for insects.} Data from more than 300 species are extracted from Chown et al \cite{Chown}. {Metabolic rates were measured for external temperatures controlled between 20 and 30ºC depending on the species.} Dashed line corresponds to the fit to a pure power law with $\alpha\approx 0.82$. Green solid line is a fit to our model. The data are highly scattered in this case and both models are equally statistically compatible (see table I).}
\label{Fig3}
\end{figure}

\noindent{In the case of insects (ectotherms), metabolic rates were measured for external temperatures controlled between 20 and 30ºC depending on the species \cite{Chown}. The scaling is plotted in fig. \ref{Fig3},} for which we find yet another different apparent exponent, $\alpha\approx 0.82$. Eq.\ref{BMR} also gives a good fit to the whole range, although for this case data are highly scattered so it is difficult to compare the accuracy of both models. {Note that in this case the effective exponent $\alpha$ is larger than what we observed for both birds and mammals (endotherms): this is consistent with our theory as insects are ectotherms and therefore the allometric term was expected to have a smaller pre-factor as thermoregulation is not present, in good agreement with empirical findings (see table \ref{table1} where it is shown that the prefactor $b$ of the allometric term for insects is one order of magnitude smaller than for birds or mammals). Incidentally, note that an analogous formula for $k'$ could be used to estimate $f$ in these cases, provided we had empirical estimates for $T_s$ for these families. Again, finding a smaller value for the pre-factor $b$ suggests a small value for $k'$, and this can also be justified as $\Delta T$ tends to be much smaller for poikilotherms.}\\

\begin{figure}
\centering
\includegraphics[width=0.7\columnwidth, angle =-90 ]{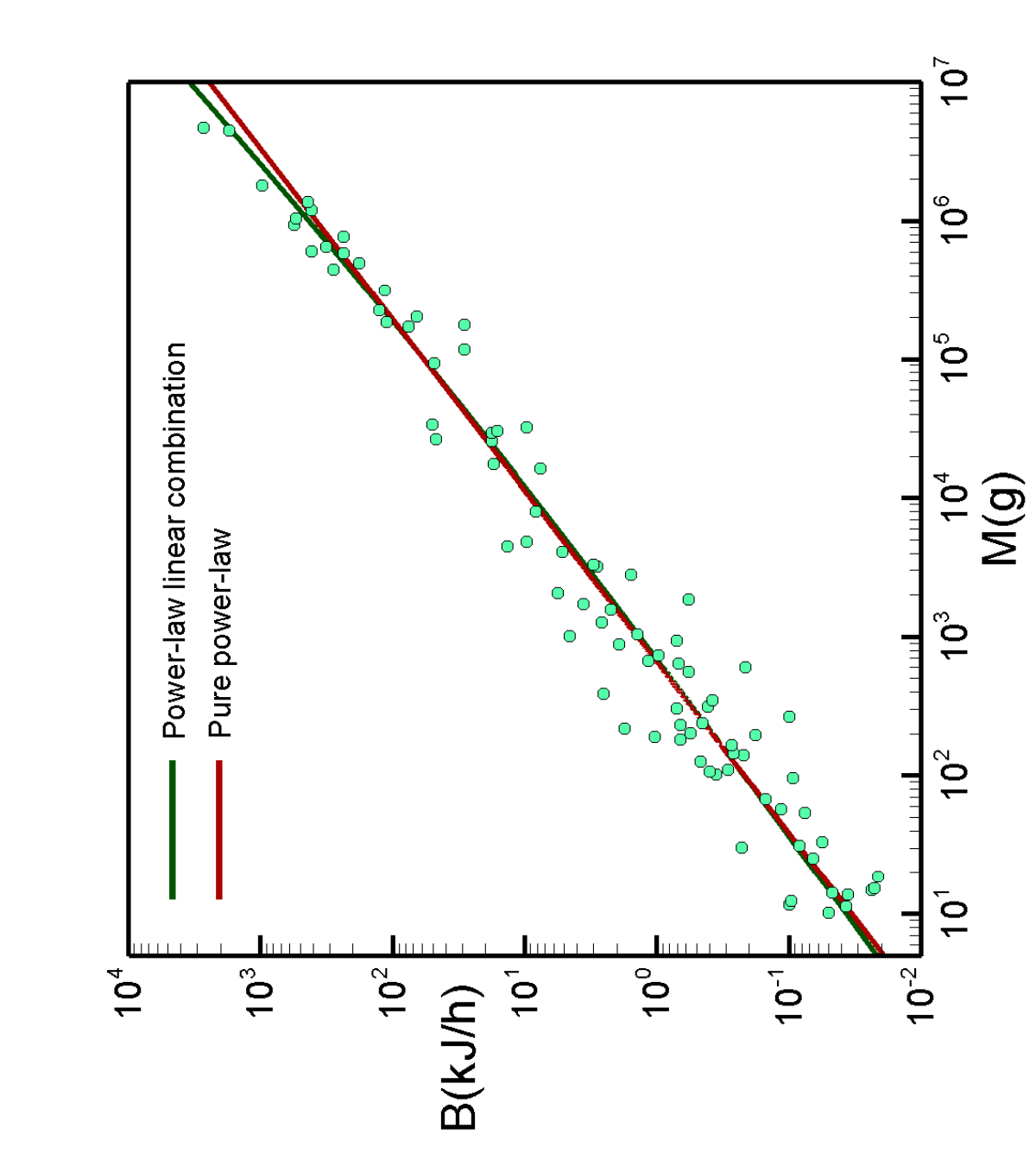}
\caption{{\bf Extension of the model for plant data.} Green dots: metabolic rate data for plants compiled by Mori et al. \cite{Mori} ($M>10$g). Green line: fitting of the model. Red line: fitting of a pure power law (see table I for statistical tests).}
\label{Fig4}
\end{figure}

\noindent {\bf Extension to plants.}
To round off, we consider the case of plants. In this case, it is neither clear what a basal metabolic rate is, nor if measurements for plants are done in their thermal neutral zone, as many are field studies in forests. Nonetheless, as plants also dissipate energy into heat our theory can be extended to this realm. Moreover, it has been found that the scaling of the respiration rate with respect to the total mass of the plant presents also a clear curvature \cite{niklas}. The term associated to heat dissipation must take into account that plants have a branched fractal surface encompassing their volume \cite{frac2}. As the surface to volume ratio is higher ($S \sim V^{D/3}$ where $2<D<3$ is the surface fractal dimension), the risk of overheating is smaller, allowing much bigger sizes than in animals. According to West et al. \cite{West} $S \sim V^{3/4}$, yielding $D \approx 2.25$ (see also \cite{zeide}), and thus our effective model reduces to $B=a M + b M^{3/4}$. As the exponents of the isometric and allometric parts are now closer, we expect a much less curved relationship with a higher effective slope ranging between 0.75 and 1. To test these predictions we have used the database of basal metabolic rates compiled by Mori et al. \cite{Mori} that includes about 200 trees and seedlings. They showed measures of metabolic rate against both total mass (including the roots) and aboveground mass. To make the comparison with mammal data homogeneous, we have used metabolic rate against total mass. Figure \ref{Fig4} shows these data, together with a fit to the model. As can be seen in table I, the fit is excellent. A pure power low model with exponent $\alpha\approx0.81$ (larger than for mammals) is a good fit as well, although our model seems to reproduce slightly better the high mass regime. As in the case of mammals, it is interesting to stress that if the second exponent of the model is left as a variable, the best fitting correctly yields the value 3/4.
As a final comment, Mori et al. stated in their paper that they found a concave curvature in their data. But this was due to the fact that they were mixing measures from adult individuals with measures from seedlings, which are growing quickly and have an altered metabolism. If we exclude seedlings and consider masses higher than 10 g we find no vestige of concavity (if fact, for masses higher that 0.1 g curvature is imperceptible).

\subsection*{Statistical methods}
Here we summarise some statistical procedures for the model fits reported in table \ref{table1}. First, we found that the dispersion of the data is multiplicative (proportional to the magnitude) and log-normally distributed. That means that error is normally distributed in logarithmic space. 
Since least-squares minimization requires errors to be normally distributed, the fitting procedure of each model consists in applying least-squares minimization to the logarithm of data (log($B$) vs. log($M$)). In other words, the nonlinear regression procedure to fit the data tries to find those values of the parameter estimates which minimize the Residual Square Error (RSS) in logarithmic space RSS$=\sum_{i=1}^N (\log(y_{obs})-\log(y_{th}))^2$.\\
Goodness of fit results include the coefficient of determination $r^2$, reduced $\chi^2$ and Akaike Information Criterion \cite{Akaike}. The  $\chi^2$ test is performed using the version of the $\chi^2$ statistic which is common in particle physics and astronomy, namely $\chi^2=\sum_{k=1}^N [y_{obs}(k)-y_{th}(k)]^2/\sigma^2$, where residuals are normalised by the standard deviation of data \cite{Bevington}. The computation of the $\chi^2$ statistic is also performed in logarithmic space. For good agreement and good estimation of standard deviations, its expected value  $\langle \chi^2 \rangle = N - p$, where $N$ is the total number of data and $p$ the number of parameters to fit. We used the reduced version  $\chi_r^2=\chi^2 /(N-p)$ whose expectation is one (i.e. good models and fits get values close to 1, and the golden rule is the smaller the better). For the Akaike Information Criterion (AIC), the golden rule is the smaller the better. AICs are computed by transforming the data into logarithmic space (where error is normally distributed) and exploiting the relation between log-likelihood and RSS via $AIC:=2k+N\ln(RSS/N)$, where $k$ is the number of free parameters in the model and $N$ the number of data points (note that for practical reasons we assume $\sigma$ to be constant across models as all have similar error distributions and thus the AIC is defined up to a constant, enabling its use to compare different models with respect to the same dataset, but not across different datasets) .

\section*{Conclusion}
 In this contribution we have built on Swan's \cite{Swan} (essential energesis is not enough to keep mammals warm) and the heat dissipation limit (HDL) paradigm \cite{Speakman} (which assumes that the capacity to dissipate heat is in fact a limit more restrictive than the energy supply). Our effective model for the body mass dependence of basal metabolic rate {was already suggested by Yates in the context of a comparison between homeotherms and poikilotherms \cite{Yates} (whose qualitative shape was already known to interpolate among several possible effective exponents \cite{White2}); here we provide a simple yet sound thermodynamic interpretation of the isometric and allometric terms according to which the model is not anymore just a fitting function. According to such interpretation, pre-factors have a physical meaning and can be measured experimentally; in this sense the model  generates self-consistent predictions which successfully account in a simple and quantitative way for a range of biologically relevant features.}\\
 
\noindent Following Glazier's proposal \cite{isometric} and the DEB predictions \cite{Kooijman} that point to combinations of isometric and allometric mechanisms operating underneath as explanation of the basal metabolic rate functional shape, we have confirmed that the wealth of different apparent exponents found for mammals, birds, insects and plants emerge possibly due to such an additive model. {Under this interpretation of Yates' model, the (now physically observable) pre-factors can vary according to exogenous conditions,} what ultimately leads to different {\it effective} exponents from a pure scaling (single power law) point of view. That is, parameters of Equation \ref{BMR} are not just fitting constants but, much on the contrary, have a physical meaning and can be empirically estimated and self-consistently predicted, as we have shown. In the case of mammals, our proposal predicts an average value $f\approx0.21$ that is confirmed by independent experimental evidence, although further measurements of direct calorimetry (heat production) and indirect calorimetry (oxygen consumption) are needed to further test this prediction. The relation between energy lost as heat and mass, the energy conversion efficiency of the metabolism, the precise curvature observed in basal metabolic rate data and their asymptotic limit, the clustering in the data between desert and polar mammals and the lower efficiency for the second group, and the higher effective exponent found in insects and plants are also results that can be explained in the light of this approach. We humbly hope that these findings can help reconciling different empirical evidence and models, and sheds some light on the role that evolutionary trade-offs between the energy dissipated as heat and the energy efficiently used by the organism to keep it alive, might play in the onset of metabolic scaling laws. {Finally, possible couplings and feedbacks at the ecological community level --due to competition and other evolutionary gradients-- might have also played a role in the ultimate shaping of these metabolic laws, which according to previous evidence have shifted across evolutionary transitions \cite{shift}. In this sense, the eventual influence from the macroecological level downwards is an open question that deserves further investigations.}


\begin{thebibliography}{100}
\bibitem{seminal} Robiquet, T. Rapport sur un m\'emoire address\'e al?Acad\'emie Royale de M\'edecin par MM Sarrus et Rameaux. {\it Bull Acad R Med Belg} {\bf 3}, 1094-1100 (1839).
\bibitem{seminal2} Rubner, M. Uber den einfluss der korpergrosse auf stoffund kraftwechsel. \textit{Z. Biol.} \textbf{19}, 536-562 (1883).
\bibitem{Krogh} { Krogh, A. {\it The Respiratory Exchange of Animals and Man}  ( Longmans, Green: London, UK, 1916).} 
\bibitem{Kleiber}	Kleiber, M. Body size and metabolism. {\it Hilgardia} {\bf 6}, 315-353 (1932)
 \bibitem{Heusner} Heusner, A. Energy metabolism and body size. I. Is the 0.75 mass exponent of Kleiber's equation a statistical artifact? {\it Respir. Physiol.}{\bf 48}, 1-12 (1982).
\bibitem{Speakman}	Speakman, J. R. $\&$ Krol, E. Maximal heat dissipation capacity and hyperthermia risk: neglected key factors in the ecology of endotherms. {\it Journal of Animal Ecology} {\bf 79}, 726-746 (2010).
\bibitem{White}	White, C. R. $\&$ Seymour, R. S. Mammalian basal metabolic rate is proportional to body mass$^{2/3}$. {\it Proc. Natl Acad. Sci. USA} {\bf 100}, 4046-4049 (2003).
\bibitem{Bartels}	Bartels, H. Metabolic rate of mammals equals the 0.75 power of their body weight. {\it Exp. Biol. Med.} {\bf 7}, 1 (1982).
\bibitem{Feldman} Feldman, H. A. $\&$ McMahon, T. A. The 3/4 mass exponent for energy metabolism is not a statistical artifact. {\it Respir. Physiol.} {\bf 52}, 149-163 (1983).
\bibitem{Savage}	Savage, V. M. et al. The predominance of quarter-power scaling in biology. {\it Funct. Ecol.} {\bf 18}, 257-282 (2004).
\bibitem{review2005} White C.R. $\&$ Seymour R.S. Allometric scaling of mammalian metabolism, \textit{J. Exp. Biol.} {\bf 208}, 1611-1619 (2005).
\bibitem{West}	West, G. B., Brown, J. H.$\&$ Enquist, B. J. A general model for the origin of allometric scaling laws in biology. {\it Science} {\bf 27}6, 122-126 (1997).
\bibitem{West2} West, G. B., Brown, J. H.$\&$ Enquist, B. J. A general model for the structure and allometry of plant vascular systems, \textit{Nature} \textbf{399} (6745):664-667 (1999).
\bibitem{Banavar2} Banavar J.R., Maritan A. $\&$ Rinaldo A. Size and form in efficient transportation networks, \textit{Nature} \textbf{399}, 130 (1999).
\bibitem{Banavar} Banavar J.R., Cooke T.J., Rinaldo A. $\&$ Maritan A. Form, function and evolution of living organisms, \textit{Proc. Natl. Acad. Sci. USA} \textbf{4}, 9 (2013).
\bibitem{Dodds}	Dodds, P. S., Rothman, D. H. $\&$ Weitz, J. S. Re-examination of the 3/4 law of metabolism. {\it J. Theor. Biol.} {\bf 209}, 9-27 (2001).
\bibitem{Kolokotrones}	Kolokotrones, T, Savage, V. Deeds, E. J. $\&$ Fontana, W. Curvature in metabolic scaling. {\it Nature} {\bf 464}, 753-756 (2010).
\bibitem{Hayssen} {Hayssen, V. $\&$ Lacy, R.C. Basal metabolic rates in mammals: Taxonomic differences in the allometry of BMR and body mass. {\it Comp. Biochem. Physiol. A Mol. Integr. Physiol.} {\bf 81} 741-754 (1985).} 
\bibitem{mackay} MacKay, N.J. Mass scale and curvature in metabolic scaling
Comment on: T. Kolokotrones et al., Curvature in metabolic scaling,  \textit{J. Theor. Biol.} \textbf{280}, 194-196 (2011).
\bibitem{clarke} Clarke, A., Rothery, P., $\&$ Isaac, N.J.B. Scaling of basal metabolic rate with body mass and temperature in mammals, \textit{J. of Animal Ecology} \textbf{79}, 610-619 (2010).
\bibitem{Hulbert14} Hubert, A.J. A Skeptics view: ``Kleiber's law" or the ``3/4 Rule" is neither a Law nor a Rule but Rather an Empirical Approximation, \textit{Systems} \textbf{2}, 186-202 (2010).
\bibitem{Glazier14} Glazier, D.S. Scaling of Metabolic Scaling within Physical Limits, \textit{Systems} \textbf{2}, 425-450 (2014).
\bibitem{PLOS} Newberry M.G., Ennis D.B. $\&$ Savage V.M. Testing Foundations of Biological Scaling Theory Using Automated Measurements of Vascular Networks, {\it PLoS Comput. Biol.} {\bf 11} (8) (2015).
\bibitem{isometric} Glazier, D.S. Beyond the "3/4-power law: Variation in the intra and interspecific scaling of metabolic rate in animals. \textit{Biol. Rev.} \textbf{80}, 611-662 (2005).
\bibitem{fito}{Mara\~non. E. Cell Size as a Key Determinant of Phytoplankton Metabolism and Community Structure, {\it Ann. Rev. Marine Science} {\bf 7} 241-264 (2015).}
\bibitem{Kozlowski}	Kozlowski, J. $\&$ Konarzewski, M. West, Brown and Enquist's model of allometric scaling again: the same questions remain. {\it Funct. Ecol.} {\bf 19} (4) 739-743 (2005).
\bibitem{McNab}	McNab, B.K. An analysis of the factors that influence the level and scaling of mammalian BMR. {\it Comp. Biochem. Physiol. A} {\bf 151}, 5-28 (2008). 
\bibitem{White2}	White, C. R. $\&$  Kearney, M. R. Determinants of inter-specific variation in basal metabolic rate. {\it J Comp Physiol B} {\bf 183}, 1-26 (2013).
\bibitem{Meeh}	Meeh, K. Oberflachenmessungen des menschlichen K\"orpers. \textit{Zeitschrift f\"ur Biologie} \textbf{15}, 425-458 (1879).
\bibitem{Glazier2014} Glazier D.D. Metabolic scaling in complex living systems, \textit{Systems} \textbf{2}, 4 (2014).
\bibitem{McNab2009} McNab, B.K. Ecological factors affect the level and scaling of avian BMR, {\it Comp. Biochem. Physiol. A} {\bf 152} 22-45 (2009).
\bibitem{Chown} Chown, S.L., Marais E., Terblanche J.S., Klok C.J., Lighton J.R.B., $\&$ Blackburn, T.M. Scaling of insect metabolic rate is inconsistent with the nutrient supply network model. {\it Func. Ecol.} {\bf 21} 282-290 (2007).
\bibitem{Blaxter}	Blaxter, K. M. {\it Energy metabolism in animals and man} (Cambridge University Press, 1989). 
\bibitem{Mortola}	Mortola, J. P. Thermographic analysis of body surface temperature of mammals. {\it Zoological Science} {\bf 30} 2, 118-124 (2013).
\bibitem{Chen} Chen, J., Patrick, R.C., Christopher, P.B., $\&$ James, D.Y.  Regulation of mitochondrial respiratory chain biogenesis by estrogens/estrogen receptors and physiological, pathological and pharmacological implications. {\it Biochimica et Biophysica Acta (BBA) - Molecular Cell Research} {\bf 1793}, 10, 1540-1570 (2009).
\bibitem{Rhoades}	Rhoades, R. A. $\&$ Bell, D. F. {\it Medical Physiology} (Lippincott Williams $\&$ Wilkins, 2013).
\bibitem{Barasi}	Barasi, M. {\it Human Nutrition: A Health Perspective} (Hodder Arnold, 2003).	
\bibitem{Sousa} Sousa, T., Mota, R., Domingos, T., $\&$ Kooijnan, S.A.L.M. Thermodynamics of organisms in the context of dynamic energy budget theory. \textit{Phys. Rev. E} \textbf{74}, 051901 (2006).
\bibitem{Sousa08} Sousa, R., Domingos, T. $\&$ Kooijnan, S.A.L.M.,From empirical patterns to theory: a formal metabolic theory of life. \textit{Phil. Trans. R. Soc. B} \textbf{363}, 2453 (2008).
\bibitem{Sousa10} Sousa, R., Domingos, T., Poggiale, J.-C. $\&$ Kooijnan, S.A.L.M. Dynamic energy budget theory restores coherence in biology. \textit{Phil. Trans. R. Soc B} \textbf{365}, 3413-3428 (2010).
  \bibitem{Nisbet12} Nisbet, R.M., Jusup, M., Klanjscek, T $\&$ Pecquerie, L. Integrating dynamyc energy budget (DEB) theory with traditional bioenergetic models.
  \textit{J. Exper. Bio.} \textbf{215}, 892-902 (2012).
\bibitem{Lodish}	Lodish, H. F. et al. {\it Molecular Cell Biology} (W. H. Freeman, 2000).
\bibitem{outliers1}  Withers, P.C.. Energy, Water, and Solute Balance of the Ostrich Struthio camelus, {\it Physiological Zoology} {\bf 56}, 4 (1983).
\bibitem{outliers2}  Maloney, S. K. $\&$ Dawson, T.J. Sexual Dimorphism in Basal Metabolism and Body Temperature of a Large Bird, the Emu. {\it The Condor} 95, 4 (1993).
\bibitem{zeide} Zeide, B. $\&$ Pfeifer. A Method for Estimation of Fractal Dimensions of Tree Crowns. \textit{Forest Science} \textbf{37}, 1253-1261 (1991).
\bibitem{Mori}	Mori S. et al. Mixed-power scaling of whole-plant respiration from seedlings to giant trees. {\it Proc. Natl Acad. Sci. USA} {\bf 107}, 1447-1451 (2010).
\bibitem{Agutter}	Agutter, P. S. $\&$ Wheatley, D. N. Metabolic scaling: consensus or controversy? {\it Theoretical Biology and Medical Modelling} {\bf 1}, 13 (2004).
\bibitem{Glazier} Glazier, D. S. A unifying explanation for diverse metabolic scaling in animals and plants. {\it Biol. Rev.} {\bf 85}, 111-138  (2010). 
\bibitem{Darveau}	Darveau, C.-A. et al.. Allometric cascade as a unifying principle of body mass effects on metabolism. {\it Nature} {\bf 417}, 166-170 (2002). 
\bibitem{Painter}	Painter, P. R. Data from necropsy studies and in vitro tissue studies lead to a model for allometric scaling of basal metabolic rate. {\it Theor. Biol. Med. Model.} {\bf 2}, 39 (2005). 
\bibitem{Kooijman} Kooijman, S.A.L.M. {\it Dynamic Energy Budget theory for metabolic organisation} (Walter de Gruyter Inc, 1990).
\bibitem{Lovegrove00} Lovegrove, B.G.  The Zoogeography of Mammalian Basal Metabolic Rate {\it The American Naturalist} {\bf 156}, 201 (2000).
\bibitem{Lovegrove03} Lovegrove, B.G.  The influence of climate on the basal metabolic rate of small mammals: a slow-fast metabolic continuum {\it J. Comp. Physiol. B} {\bf 173}, 87-112 (2003).
\bibitem{fristoe} Fristoe, T.S., Burger, J.R., Balk, M.A., Khaliq, I., Hof, C. $\&$ Brown, J.H., Metabolic heat production and thermal conductance are mass-independent adaptations to thermal environment in birds and mammals. {\it Proc. Natl. Acad. Sci. USA}  {\bf 112}, 15934-15939 (2015).
\bibitem{Zotin} Zotin, A.I. {\it Thermodynamic Bases of Biological Processes} (Cambridge Univ. Press, 2010).
\bibitem{K} Kozlowski, J., Konarzewski, M $\&$ Gawelczyk, A.T. Cell size as a link between noncoding DNA and metabolic rate scaling. {\it Proc. Natl. Acad. Sci. USA} {\bf 100}, 24 (2003).
\bibitem{Economos}	Economos, A. C. Gravity, metabolic rate and body size of mammals. {\it Physiologist} {\bf 22} S71 (1979).
\bibitem{Swan}	Swan, H. {\it Thermoregulation and bioenergetics: patterns for vertebrate survival} (American Elsevier, 1974).
\bibitem{Bevington}	Bevington, P. R. $\&$ Robinson, D. K. {\it Data reduction and error analysis for the physical sciences} (Mc Graw Hill, 2003).
\bibitem{Akaike}	Akaike, H. A new look at the statistical model identification. {\it IEEE Transactions on Automatic Control} {\bf 19} (6) 716-723 (1974).
\bibitem{niklas} Niklas, K.J., $\&$ Kutschera, U. Kleiber's Law: How the Fire of Life ignited debate, fueled theory, and neglected plants as model organisms \textit{Plant Sig. \& Behavior} \textbf{10} e1036216 (2015).
\bibitem{frac2} Rodriguez-Iturbe, I. $\&$ Rinaldo, A. {\it Fractal River Basins: Chance and Self-Organization} (Cambridge University Press, 1997).
\bibitem{Yates}{Yates F.E. Comparative physiology of energy production: homeotherms and poikilotherms. {\it Am. J. Physiol.-Reg., Integr. Comp. Physiol.} {\bf 240} R1-R2 (1981).}
\bibitem{shift} {DeLong, J.P., Okie, J.G., Moses, M.E., Sibly, R.M. $\&$ Brown, J.H. Shifts in metabolic scaling, production, and efficiency across major evolutionary transitions of life. {\it Proc. Natl. Acad. Sci. USA} {\bf 107} (2010).} 
\bibitem{iain}{J. Aryaman, H. Hoitzing, J.P. Burgstaller, I.G. Johnston $\&$ N.S. Jones, Mitochondrial heterogeneity, metabolic scaling and cell death, {\it Bioessays} {\bf 39}, 7 (2017).}
\end{thebibliography}

\section*{Acknowledgements}

We thank D. Martinez for his help in the selection of hot desert animals, A. Robledo, J. Cuesta, A. Fernandez and O. Miramontes for their useful comments {and anonymous referees for very insightful remarks.}

\section*{Author contributions statement}

FJB and VJM conceived the initial theory. FJB, VJM, BL, LL, AM and EV developed the final theory. FJB, BL and LL analysed and fitted the data. All authors reviewed the manuscript. This work has been funded by projects AYA2013-48623-C2-2, FIS2013-41057-P, CGL2013-46862-C2-1-P and SAF2015-65878-R from the Spanish Ministerio de Econom\'{\i}a y Competitividad and 
PrometeoII/2014/086, PrometeoII/2014/060 and PrometeoII/2014/065 from the Generalitat Valenciana (Spain). BL acknowledges funding from a Salvador de Madariaga fellowship, and LL acknowledges funding from EPSRC Early Career fellowship EP/P01660X/1.

\section*{Competing financial interests}

The authors have no competing interests as defined by Nature Publishing Group, or other interests that might be perceived to influence the results and/or discussion reported in this paper.

\begin{table*}
\begin{tabular}{|llcccc|}
\hline 
 {\bf Database $\&$ model} & {\bf Parameters fit}& $b/a$  & $r^2 (\%)$ &  $\chi_r^2 $ &AIC \\
\hline
{\it Mammals (all, $N=637$)}&&&&&\\
$B= c M^{\alpha}$ & $c=0.0692$, $\alpha=0.72$ &  &95.2 & $1.07$ & -1220\\
$\log B = \beta_0 + \beta_1 \log M + \beta_2 (\log M)^2$ & $\beta_0=-2.19$, $\beta_1=0.54$, $\beta_2=0.014$ & &96.1& $1.01$&-1271\\
$B=aM+bM^{2/3}$ & $a=0.0016$, $b=0.079$& 49&97.9&  $1.00$&-1264 \\
\hline
{\it Polar mammals ($N=14$)}&&&&&\\
$B= c M^{\alpha}$ & $c=0.1326$, $\alpha=0.6928$ &  & 98.8& $0.86$ & -32.0\\
$B=aM+bM^{2/3}$ & $a=0.00085$, $b=0.142$& 167&98.9&  $0.78$& -33.2 \\
\hline
{\it Desert mammals ($N=99$)}&&&&&\\
$B= c M^{\alpha}$ & $c=0.0556$, $\alpha=0.7393$ &  & 96.9&  $0.94$& -233\\
$B=aM+bM^{2/3}$ & $a=0.002$, $b=0.066$& 29&97.2&  $0.93$& -244 \\
\hline
{\it Polar and Desert mammals ($N=113$)}&&&&&\\
$B= c M^{\alpha}$ & $c=0.0569$, $\alpha=0.7468$ &  & 96.9& 1.02 & -233\\
$B=aM+bM^{2/3}$ & $a=0.002$, $b=0.0706$& 35&97.0&  0.97& -245 \\
Hybrid&&&98.0& 0.90& -277\\
\hline
{\it Plants ($N=89$)}&&&&&\\
$B= c M^{\alpha}$ & $c=0.0053$, $\alpha=0.81$ & &95.7&  $1.03$ & -77.5\\
$B=aM+bM^{3/4}$ & $a=0.00021$, $b=0.0064$& 30&95.8&  $1.00$& -78.7 \\
$B=aM+bM^{\beta}$ & $a=0.00021$, $b=0.0064$, $\beta=0.750$& 30&95.8&  $1.00$& -76.7 \\
\hline
{\it Flying Birds ($N=510$)}&&&&&\\
$B= c M^{\alpha}$ & $c=0.143$, $\alpha=0.657$ & &88.4& 1.04 &-1309 \\
$B=aM+bM^{2/3}$ & $a=0.0001$, $b=0.137 $& 1370&90.9&  1.01 & -1308 \\
\hline
{\it Flightless Birds (all, $N=22$)}&&&&&\\
$B= c M^{\alpha}$ & $c=0.062$, $\alpha=0.744$ & &90.2&  0.75 &-55.0 \\
$B=aM+bM^{2/3}$ & $a=0.0014$, $b=0.092 $& 66&85.7&  0.69 &-52.8 \\
\hline
{\it Flightless Birds (without outliers, $N=20$)}&&&&&\\
$B= c M^{\alpha}$ & $c=0.041$, $\alpha=0.805$ & & 98.6& 0.88 & -53.5\\
$B=aM+bM^{2/3}$ & $a=0.0042$, $b=0.062$& 17&98.7&  0.88 & -54.5 \\
\hline
{\it Insects}&&&&&\\
$B= c M^{\alpha}$ & $c=0.007$, $\alpha=0.832$ & &60.4&  1.04  & 0.85 \\
$B=aM+bM^{2/3}$ & $a=0.0046$, $b=0.0021$& 0.46 &58.9&  1.09 & 7.00 \\
\hline
\end{tabular}
\caption{\label{table1} {\bf Fitting results and goodness of fit}. See Statistical Methods section for details. The first three columns present the different fitting models considered for the different datasets, along with the parameter fits and ratio $b/a$ (when applicable). 
The following columns display the goodness of fit results: the coefficient of determination $r^2$, reduced $\chi^2$ and Akaike Information Criterion \cite{Akaike}. 
In every case, we find that Eq. \ref{BMR} is statistically compatible with the data and has in several cases better goodness of fit than other fitting models. A model selection approach (based on AIC) suggests that Eq. \ref{BMR} outperforms a pure power law model with varying exponent for mammals, polar mammals alone, desert mammals alone, polar and desert mammals alone, flightless birds without outliers and plants. Additionally, note that the pure power law fitting model systematically requests different power law exponents  for different databases, challenging the validity of the $2/3$ or $3/4$ laws, whereas in Eq. \ref{BMR} the exponents are fixed and only prefactors vary.}
\end{table*}

\end{document}